\documentclass[prd,twocolumn,reprint,preprintnumbers,nofootinbib]{revtex4-1}

\input{header}

\begin{document}
\title{FORESEE: FORward Experiment SEnsitivity Estimator\\ for the LHC and future hadron colliders}

\author{Felix Kling}
\email{felixk@slac.stanford.edu}
\affiliation{Theory Group, SLAC National Accelerator Laboratory, Menlo Park, California 94025}

\author{Sebastian Trojanowski}
\email{strojanowski@camk.edu.pl}
\affiliation{Astrocent, Nicolaus Copernicus Astronomical Center Polish Academy of Sciences, ul.~Rektorska 4, 00-614, Warsaw, Poland}
\affiliation{National Centre for Nuclear Research, ul.~Pasteura 7, 02-093 Warsaw, Poland
}

\begin{abstract}
We introduce a numerical package \textbf{FOR}ward \textbf{E}xperiment \textbf{SE}nsitivity \textbf{E}stimator, or \texttt{FORESEE}, that can be used to simulate the expected sensitivity reach of experiments placed in the far-forward direction from the proton-proton interaction point. The simulations can be performed for $14~\tev$ collision energy characteristic for the LHC, as well as for larger energies: $27$ and $100~\tev$. In the package, a comprehensive list of validated forward spectra of various SM species is also provided. The capabilities of \texttt{FORESEE} are illustrated for the popular dark photon and dark Higgs boson models, as well as for the search for light up-philic scalars. For the dark photon portal, we also comment on the complementarity between such searches and dark matter direct detection bounds. Additionally, for the first time, we discuss the prospects for the LLP searches in the proposed future hadron colliders: High-Energy LHC (HE-LHC), Super proton-proton Collider (SppC), and Future Circular Collider (FCC-hh).
\end{abstract}

\maketitle

\section{Introduction}
\label{sec:intro}

The Large Hadron Collider (LHC) has proven to be a powerful tool for studying both new physics and the Standard Model (SM), with its most remarkable achievement related to the discovery of the Higgs boson~\cite{Chatrchyan:2012ufa,Aad:2012tfa}. While further such outstanding signatures of new particles are much awaited, the data collected in the LHC have also already been used to constrain many beyond the Standard Model (BSM) scenarios in a tremendous number of phenomenological analyses. This became feasible due to the development of convenient modeling tools that were made available to a wider community. 

This relentless quest for new physics will not only be continued in the upcoming LHC Run 3, but it will also be further extended to better encompass growing interest in searches for new light and long-lived particles (LLPs), cf. Refs~\cite{Beacham:2019nyx,Agrawal:2021dbo} for recent reviews. While the traditional BSM physics programs target new particles with masses typically above $\mathcal{O}(10~\gev)$ up to a few \tev\ scale, lighter new physics species could become more accessible with modified experimental strategies~\cite{Alimena:2019zri}. In particular, this observation has lead to the establishment of a new direction in the BSM searches at the LHC in its far-forward region, as originally proposed~\cite{Feng:2017uoz,Feng:2017vli,Kling:2018wct,Feng:2018noy} for the FASER experiment~\cite{Ariga:2018zuc,Ariga:2018uku,Ariga:2018pin}. Further such opportunities could be explored in the high-luminosity LHC (HL-LHC) era with an expanded research agenda of the Forward Physics Facility (FPF)~\cite{SnowmassFPF}. 

In order to fully exploit the relevant physics potential, the experimental efforts should be supplemented with a comprehensive program of theoretical and phenomenological studies. To facilitate this, we introduce a numerical package, namely the \textbf{FOR}ward \textbf{E}xperiment \textbf{SE}nsitivity \textbf{E}stimator, or \texttt{FORESEE}, which could be used to obtain the expected sensitivity reach for BSM models in various far-forward experiments.\footnote{available at \href{https://github.com/KlingFelix/FORESEE}{https://github.com/KlingFelix/FORESEE}}

The package allows one to perform quick but accurate simulations for selected popular BSM simplified models. This can be done for user-defined experimental geometries and basic cuts applied to the visible signal. The package also provides a set of useful numerical data, including e.g. the far-forward spectra of light mesons, that can easily be accessed and employed in separate studies estimating the new physics sensitivity reach in other BSM scenarios. We illustrate below the capabilities of \texttt{FORESEE} for the popular dark photon and dark Higgs boson models, as well as for the model with a hadrophilic dark scalar with the dominant couplings to the up quarks~\cite{Batell:2017kty,Batell:2018fqo}. In addition, for the first time, we study the possibility to perform such far-forward searches in the future $27~\tev$ and $100~\tev$ hadron colliders.

This study is organized as follows. In \cref{sec:foresee}, we introduce the \texttt{FORESEE} package. We briefly present the validated far-forward spectra of hadrons and other SM particles that can be used for BSM modeling. We then discuss the capabilities of the automated numerical tool that we provide in studying the sensitivity reach of far-forward experiments. In \cref{sec:physics}, the \texttt{FORESEE} is employed to analyze specific examples of the far-forward LLP searches both in the LHC and future hadron colliders. We conclude in \cref{sec:conclusion}.

\begin{figure*}[!ht]
\centering
\includegraphics[width=0.48\textwidth]{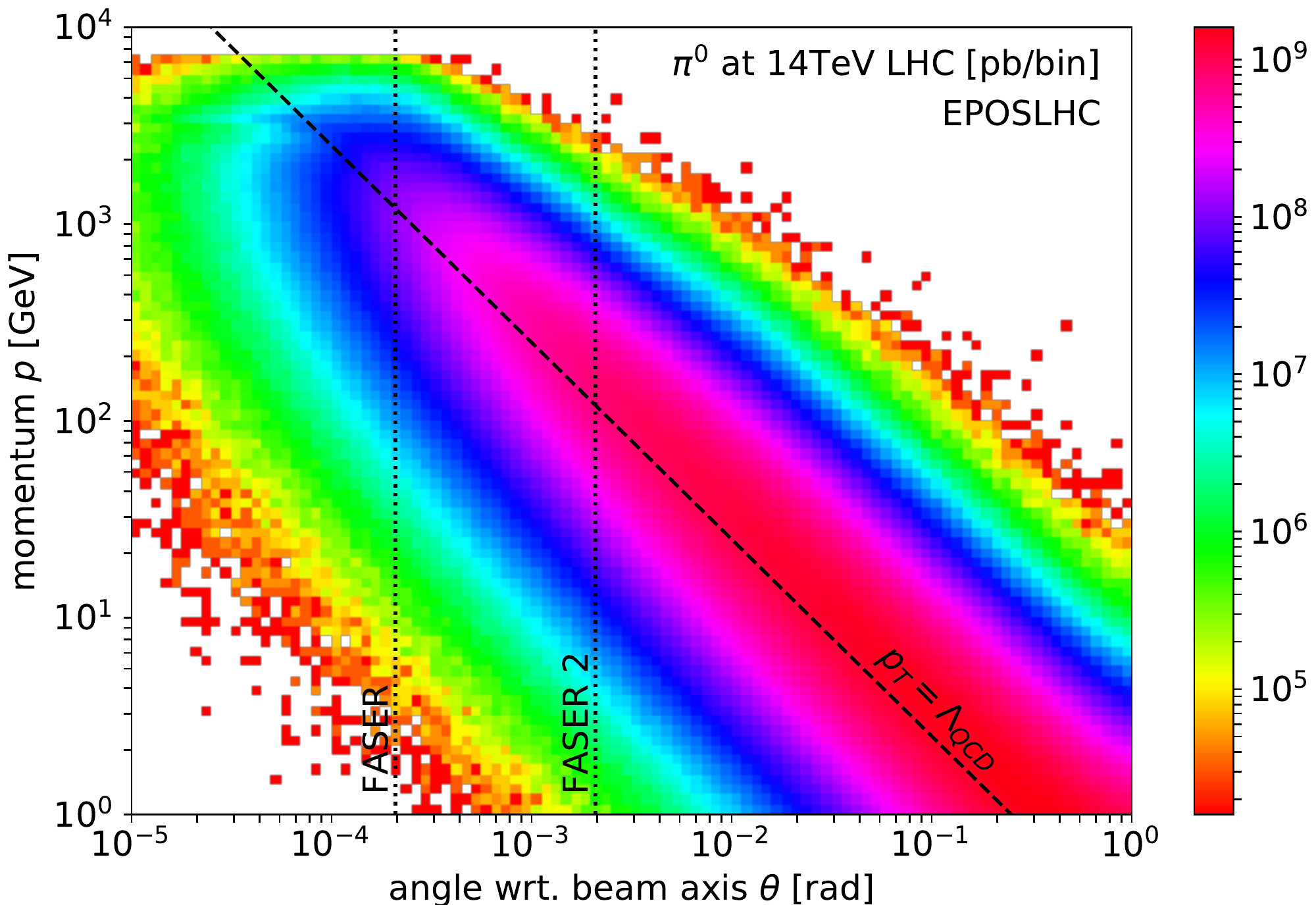}
\hspace{3mm}
\includegraphics[width=0.48\textwidth]{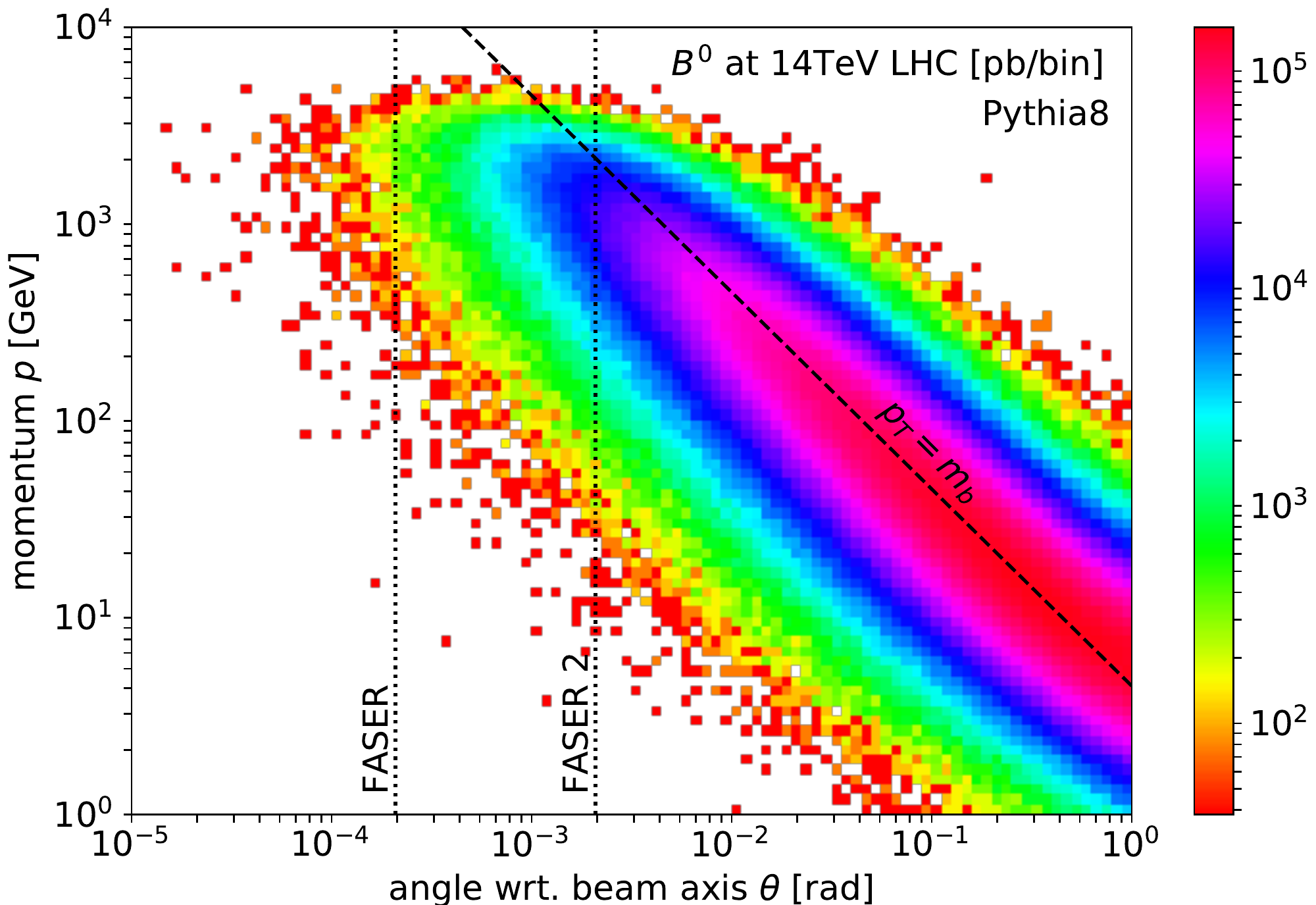}%
\caption{Distribution of $\pi^0$ (left) and $B^0$ (right) mesons in the forward hemisphere in the ($\theta$,$p$) plane, where $\theta$ and $p$ are the meson’s angle with respect to the beam axis and momentum, respectively. We use $20$ bins per decade and present the spectra obtained for $14~\tev$ $pp$ collision energy. The $\pi^0$ spectrum is obtained with EPOS-LHC~\cite{Pierog:2013ria}, while the $B$-meson one with \texttt{Pythia}~\cite{Sjostrand:2014zea} with the Monash tune~\cite{Skands:2014pea}. The diagonal black dashed lines highlight the characteristic transverse momentum scale $p_T\sim\Lambda_{\textrm{QCD}}\sim 250~\mev$ for pions and $p_T\sim m_B$ for $B$ mesons. The angular acceptances for the FASER and FASER 2 experiments to take data during LHC Run 3 and in the HL-LHC era, respectively, are indicated by the vertical black dotted lines.}
\label{fig:spectra}
\end{figure*}

\section{FORESEE package}
\label{sec:foresee} 

\subsection{Forward spectra of SM particles}
\label{sec:hadrons} 

Light new physics particles can be efficiently produced in the far-forward region of the LHC due to at least several different production mechanisms, cf. Refs~\cite{Ariga:2018uku,Jodlowski:2019ycu,Jodlowski:2020vhr} for further discussion. A prominent role among them is played by rare decays of light mesons that are abundantly created in $pp$ collisions. Notably, one expects e.g. about $N_{\pi^0}\sim 4 \times 10^{17}$ neutral pions to be produced during Run 3. In particular, high-energy mesons typically travel far-forward at angles $\theta_{M}\sim p_{T,M}/p_M\ll 1$ with respect to the beam axis. Here $p_{T,M}\sim m_{M}\sim \mathcal{O}(100~\mev-\textrm{a few }\gev)$ is the meson's characteristic transverse momentum of order the meson mass, and $p_M\sim\tev$ is the meson's total momentum. Importantly, such meson decays are typically the dominant LLP production channels as long as they are kinematically available.

Properly predicting the meson spectra is then essential for modeling new physics effects in far-forward experiments. This can be done using dedicated hadronic interaction models, designed to  describe inelastic collisions at both particle colliders and cosmic ray experiments. Those models have greatly improved in recent years, partially due to the dedicated forward physics program at the LHC~\cite{Akiba:2016ofq}. In the left panel of \cref{fig:spectra}, we present exemplary such spectra of neutral pions and $B$ mesons in the $(\theta_M,p_M)$ plane. These were obtained, respectively, with the \texttt{EPOS-LHC}~\cite{Pierog:2013ria} event generator implemented in the \texttt{CRMC} simulation package~\cite{CRMC}, and by using \texttt{Pythia~8}~\cite{Sjostrand:2006za, Sjostrand:2014zea} with the Monash tune~\cite{Skands:2014pea}. As expected, the spectra are concentrated along the $p_{T,M}\sim m_M$ lines.  

\renewcommand{\arraystretch}{1.2}
\setlength{\tabcolsep}{5pt}
\begin{table*}[t!]
\centering
    \begin{tabular}{c|c||c|c|c|c}
      \hline\hline
     \multirow{2}{*}{Particle category} & \multirow{2}{*}{Particles} & \multicolumn{4}{c}{Generators}\\
    \cline{3-6}
     & & EPOS-LHC & QGSJET II-04 & SIBYLL 2.3c & \texttt{Pythia 8}\\
     \hline
 Photons & $\gamma$ & \checkmark & \checkmark & \checkmark & \\
 \multirow{2}{*}{Light hadrons} & $\pi^0$, $\pi^+$, $\eta$, $\eta^\prime$, $\omega$, $\rho$, $\phi$, $n$, $p$ & \multirow{2}{*}{\checkmark} & \multirow{2}{*}{\checkmark} & \multirow{2}{*}{\checkmark} & \multirow{2}{*}{} \\
 & $K^+$, $K_L$, $K_S$, $K_0^*$, $K^{*+}$, $\Lambda$& & & & \\
 Charm hadrons & $D^+$, $D^0$, $D_s^+$, $\Lambda_c$ &  &  & \checkmark & \checkmark \\
 Beauty hadrons & $B^0$, $B^+$, $B_s$, $B_c^+$, $\Lambda_b$ &  &  &  & \checkmark \\
 Heavy quarks & $c$, $b$ &  &  &  & \checkmark \\
 Quarkonia & $J/\Psi$, $\psi(2S)$, $\Upsilon(1S)$, $\Upsilon(2S)$, $\Upsilon(3S)$ &  &  &  & \checkmark \\
 Weak bosons & $W^+$, $Z$, $h$ &  &  &  & \checkmark \\
 \hline \hline
 \end{tabular}
 \caption{Far-forward spectra of Standard Model particles that can be employed in simulations with \texttt{FORESEE} and the Monte Carlo simulation tools used to obtain them (see the text for references). The spectra are available in the package as text files that can be found in the \texttt{files/hadrons} directory.}
 \label{tab:spectra}
\end{table*}

In the \texttt{FORESEE} package, these spectra are then used to generate forward LLP flux due to the relevant meson decays. Since in some applications it might be beneficial to access only these spectra and employ them in independent simulations, we provide them in a simple format in separate text files for different mesons and three $pp$ collision energies: $14~\tev$, $27~\tev$, and $100~\tev$. The relevant files can be found in the package's \texttt{files/hadrons} directory. The meson spectrum is provided in tables with the consecutive columns corresponding to the bin position in ($\log_{10}{\theta_M}$, $\log_{10}{(p_M/\gev)}$) and the weights of each bin ($3$rd column) in units of pb/bin. Here $\theta_M$ in radians is the angle with respect to the beam axis. To obtain the number of mesons per bin, the bin weights should be multiplied by the relevant integrated luminosity (e.g., $150~\ifb$ for LHC Run 3, $3000~\ifb$ for HL-LHC). The number in each of the filenames is the meson ID in the MC particle numbering scheme~\cite{Zyla:2020zbs}. The production cross section is given for the forward hemisphere only. We similarly treat selected other SM species listed below.

We provide the spectra for the following light mesons: $\pi^0$, $\pi^\pm$, $K^\pm$, $K_L$, $K_S$, $K_0^*$, $K^{*\pm}$, $\eta$, $\eta^\prime$, $\omega$, $\rho$ and $\phi$, for the baryons: $n$, $p$ and $\Lambda$ and their anti-particles, as well as for photons $\gamma$. The user can access the results obtained for three different MC generators that should also be referred to when using the spectra, namely EPOSLHC~\cite{Pierog:2013ria}, QGSJET II-04~\cite{Ostapchenko:2010vb}, and SIBYLL~2.3c~\cite{Ahn:2009wx,Riehn:2015oba} as implemented in \texttt{CRMC}~\cite{CRMC}. For heavier SM species, we employ SIBYLL and \texttt{Pythia~8}~\cite{Sjostrand:2006za, Sjostrand:2014zea} with the Monash tune~\cite{Skands:2014pea} for charmed hadrons ($D^0$, $D^+$, $D_s$, $\Lambda_c$, and $c,\bar c$ quarks), while we use \texttt{Pythia~8} for bottom hadrons ($B^0$, $B^+$, $B_s$, $B_c$, $\Lambda_b$ and $b, \bar b$ quarks), heavy gauge bosons $W$ and $Z$, and for the SM Higgs boson $h$. In addition, we also provide the spectra of $J/\psi$, $\psi(2S)$, and $\Upsilon(1,2,3S)$ mesons for 14 TeV LHC, see Ref.~\cite{Foroughi-Abari:2020qar} for further discussion and validation of these spectra. In \cref{tab:spectra}, we summarize the SM spectra available in \texttt{FORESEE} and the relevant simulation tools used to obtain them.

\subsection{Sensitivity Reach Estimator}
\label{sec:code} 

On top of providing the users with the meson spectra files, the \texttt{FORESEE} package can also be used as a standalone simulation tool to study the sensitivity reach of far-forward experiments at the LHC and in future hadron colliders. The simulations can be performed for all the three aforementioned $pp$ collision energies. Below, we briefly describe the consecutive steps of such simulations. Further instructions on how to use the code are provided in the package in tutorial jupyter notebooks. 

In the first step of the simulation, the user must define the model by specifying \textsl{i)} the LLP production rates, \textsl{ii)} their lifetimes, and \textsl{iii)} the LLP decay branching fractions. The last information is optional and, if it is not defined, it is assumed that the experiment is sensitive to all the LLP decays happening inside the detector.

The \texttt{FORESEE} package supports three different production modes: SM particle decays, mixing with SM particles, and direct production. In most models, the main production modes of LLPs are the decays of SM particles. For 2-body decays and unpolarized beams, the kinematics are fully specified by the masses of the parent and daughter particles, in which case one only needs to provide the total decay branching fraction. There is also the option for 3-body decays, $p_0 \to p_1 p_2 p_3$, with $p_3$ being the LLP. Here ones needs to provide the differential branching fraction $d\text{BR}/(dq^2 \ d\cos\vartheta)$, where $q^2=(p_2+p_3)^2$ and $\vartheta$ is the angle between $p_3$ in the rest frame of $p_2+p_3$, and the direction of $p_2+p_3$ in the rest frame of $p_0$. When modeling the production of LLPs in decays of long-lived mesons (charged pions, charged/neutral kaons), we additionally require the mesons to decay before they hit the beampipe at radius $r=5~\cm$, the neutral particle absorber TAN at a distance of $z=140~\m$ from the $pp$ Interaction Point (IP), or the inner triplet quadrupole absorber TAS at $z=20$m. Charged particles are always required to decay before the first magnets at $z=20~\m$.

In addition, LLPs can also be produced via their mixing with the SM mesons. Examples are the mixing of axion-like particles with the SM pseudoscalar mesons or dark photons with the SM vector bosons, which were discussed in Ref.~\cite{Ariga:2018uku} and Ref.~\cite{Berlin:2018jbm}, respectively. Here we assume that the LLP production rate is related to the SM particle production rate via $\sigma(LLP) = \kappa^2 \times \sigma(SM)$, where $\kappa$ describes the mixing which must be provided by the user. 

Finally, LLPs can also be produced directly. Examples for this include dark photon production via Bremsstrahlung or Drell-Yan production (see Ref.~\cite{Feng:2017uoz} and Ref.~\cite{Berlin:2018jbm} for details on these production modes). In this case, the user must provide the full two-dimensional LLP spectra for different LLP masses in the same format as previously used for the SM particles.

The LLP lifetime, $c\tau$, and decay branching fractions need to be provided by the user in form of a table. The user can choose either a one-dimensional or a two-dimensional parameterization. In the first case, the lifetime is given for the specific value of the coupling constant $g_*$ as a function of the varying LLP mass $m$. For different values of the coupling constant $g$, one evaluates the lifetime as $c\tau(m,g)=c\tau(m,g_*)g_*^{2}/g^2$. In the two-dimensional parameterization, $c\tau(m,g)$ is provided directly as a function of both $m$ and $g$. In both cases, the code then interpolates the lifetime throughout the parameter space. The same options are available for the branching fractions. 

\renewcommand{\arraystretch}{1.2}
\setlength{\tabcolsep}{5pt}
\begin{table*}[t!]
\centering
    \begin{tabular}{c||c|c|c|c|c}
      \hline\hline
    Collider & Luminosity $\mathcal{L}$& Energy & Distance $L$ & Detector Length $\Delta$ & Detector Radius $R$\\
      \hline
      FASER, LHC  & $150~\ifb$ &$14~\tev$ & $480~\m$ & $1.5~\m$ & $10~\cm$ \\
      FASER 2, HL-LHC & $3000~\ifb$ &$14~\tev$ & $480~\m$ & $5~\m$ & $1~\m$ \\
      HE-LHC & $15~\iab$ &$27~\tev$ & $480~\m$ & $5~\m$ & $1~\m$ \\
      FCC/SppC & $30~\iab$ &$100~\tev$ & $1~\km$ & $5~\m$ & $1~\m$ \\
      \hline \hline
    \end{tabular}
    \caption{The details of the detectors used in the analysis. All detectors are assumed to have cylindrical geometry with the radius $R$ and the length $\Delta$. They are placed at a distance $L$ from the $pp$ interaction point and are centered along the beam collision axis.}
    \label{tab:geometries}
    \end{table*}

After specifying the model, \texttt{FORESEE} generates the two-dimensional LLP spectra in terms of $(\theta_{LLP},p_{LLP})$. For the next step of the analysis, the user can specify the \textsl{i)} distance $L$ between the IP and experiment, \textsl{ii)} acceptance in terms of the LLPs momentum and position, \textsl{iii)} luminosity, \textsl{iv)} production channels to consider, and \textsl{v)} allowed LLP decay channels. \texttt{FORESEE} then counts the number of signal events that pass the selection criteria. In particular, the package allows the user to define the detector position and geometry, as well as the analysis cuts, by restricting the allowed position and momentum $3$-vectors of the LLP. 

The reach plots are automatically generated for a user-defined number of BSM signal events. In the plots, current bounds are also shown, as well as the expected reach for selected proposed future experiments. For the detector under study, the default number of signal events, which defines the sensitivity line to be drawn, is $N_{\textrm{ev}}=3$. This appropriately estimates the reach in searches with negligible SM backgrounds. It should be stressed, however, that \texttt{FORESEE} only estimates the BSM signal rate. Depending on the detector location and design, there could be a large number of background events that could affect the searches for new physics. For example, in detectors with a material-filled decay volume, scattering signatures from neutrinos, photons, or neutral hadrons could mimic LLP decays if they are not carefully rejected in the analysis. Alternatively, locations not well shielded from the LHC beam and infrastructure could suffer from significant backgrounds from the beam debris, which would have to be taken into account in the modeling. In order to allow the user to include such effects when estimating the BSM sensitivity reach of a given experiment, we leave freedom to define the aforementioned number of signal events $N_{\textrm{ev}}$ to be presented in the plots. 

Further developments of the package are planned for the future that will add more popular LLP models, the relevant production and decay modes, as well as will expand related modeling capabilities. 

\section{Physics Examples}
\label{sec:physics} 

In this section, we illustrate the capabilities of \texttt{FORESEE} by presenting the expected sensitivity reach plots for selected specific BSM scenarios and the LLP decay signature. We first focus on the FASER and FASER 2 experiments at the LHC and then move to discuss the prospects for LLP searches in the far-forward region of the future hadronic colliders. 

When presenting our results, we assume that background in the searches for decays of high-energy LLPs, typically with $E_{\textrm{LLP}}\gtrsim 100~\gev$, can be reduced to negligible levels. The possibility of such a background reduction has been studied in detail for the FASER experiment to take data during LHC Run 3~\cite{Ariga:2018zuc,Ariga:2018pin}. In the case of other detectors under study, we expect that a similar suppression will be possible, as they would be shielded from the $pp$ collision point by at least $\mathcal{O}(100~\m)$ of the rock and other elements of the collider infrastructure.

\subsection{Models\label{sec:models}}

For illustration, we will present the results for the models predicting the existence of sub-$\gev$ dark photons $A^\prime$, dark Higgs bosons $\phi$, and the up-philic scalars $S$. The corresponding interaction Lagrangians are given by:
\begin{align}
\label{eq:LAprime}
\mathcal{L} & = \frac{1}{2}\,m_{A'}^2\,A^{\prime\,2} - \epsilon\,e\,q_f\,\bar{f}\,\slashed{A}^\prime\,f\hspace{1.1cm}(\textrm{dark photon}),\\
\label{eq:darkHiggseffL}
\mathcal{L} & = -m_\phi^{2} \phi^2 - \sin{\theta} \frac{m_f}{v} \phi \bar{f} f  
-  \lambda vh \phi\phi\,\,(\textrm{dark Higgs}),\\
\label{eq:Lupphilicscalar}
\mathcal{L} & = -m_S^{2} S^2 - g_u\,S \bar{u}_L u_R  \hspace{1.2cm} (\textrm{up-philic scalar}),
\end{align}
where $\bar{f}f$ represents the sum over all the SM fermions and $h$ is the SM Higgs boson. Above, we also define the relevant coupling constants: the kinetic mixing parameter $\epsilon$ for $A^\prime$, the mixing angle $\theta$ and trilinear coupling $\lambda$ for the dark Higgs boson, and the $g_u$ coupling for the up-philic scalar. Further details of the far-forward modeling for these scenarios can be found in Ref.~\cite{Feng:2017uoz} for $A^\prime$s and in Ref.~\cite{Feng:2017vli} for the dark Higgs boson, while the case of the up-philic scalar is described below.

These example models are characterized by different dominant production or decay modes of LLPs. In particular, light dark photons with $m_{A^\prime}\lesssim \mathcal{O}(100~\mev)$ are dominantly produced in $\pi^0$ decays, the up-philic scalars come from $\eta$ decays, and the dark Higgs bosons are mostly generated in rare decays of $B$ mesons. Notably, while they typically lead to LLP decays into two opposite-charge SM particles, the light up-philic scalars can also have the leading decay channel into two photons, cf. Refs~\cite{Feng:2018noy,Kling:2020mch} for other examples of far-forward searches for LLPs with the dominant di-photon decay channel.  

\subsection{Dark photon}
\label{sec:darkphoton} 

In the left panel of \cref{fig:LHC}, we show the expected sensitivity reach of the FASER and FASER 2 experiments in the search for the vanilla dark photons, cf. \cref{eq:LAprime}. The assumed detector geometries and the collider luminosities are given in \cref{tab:geometries}. The total sensitivity reach lines obtained with \texttt{FORESEE} and presented in the plot, reproduce the results obtained previously by the FASER collaboration~\cite{Ariga:2018uku}.

In addition, we present in the plot the expected reach of FASER 2 assuming that only the simplest $A^\prime\to e^+e^-$ decay channel is considered in the analysis. As expected, this has no impact on the reach for the dark photon mass below the di-muon threshold, while it moderately reduces the sensitivity for $m_{A^\prime}\gtrsim 500~\mev$. Here, possible $A^\prime$ decays into hadronic final states start to play the dominant role and they would have to be taken into account in the analysis. The relevant hadronic branching fraction is dominated by the decays into two charged pions, $A^\prime\to \pi^+\pi^-$, while in narrow ranges of the $A^\prime$ mass around the $\omega$ or $\phi$ resonances decays with neutral pions or kaons in the final state become also important~\cite{Buschmann:2015awa}. 

\begin{figure*}[t]
\centering
\includegraphics[width=0.49\textwidth]{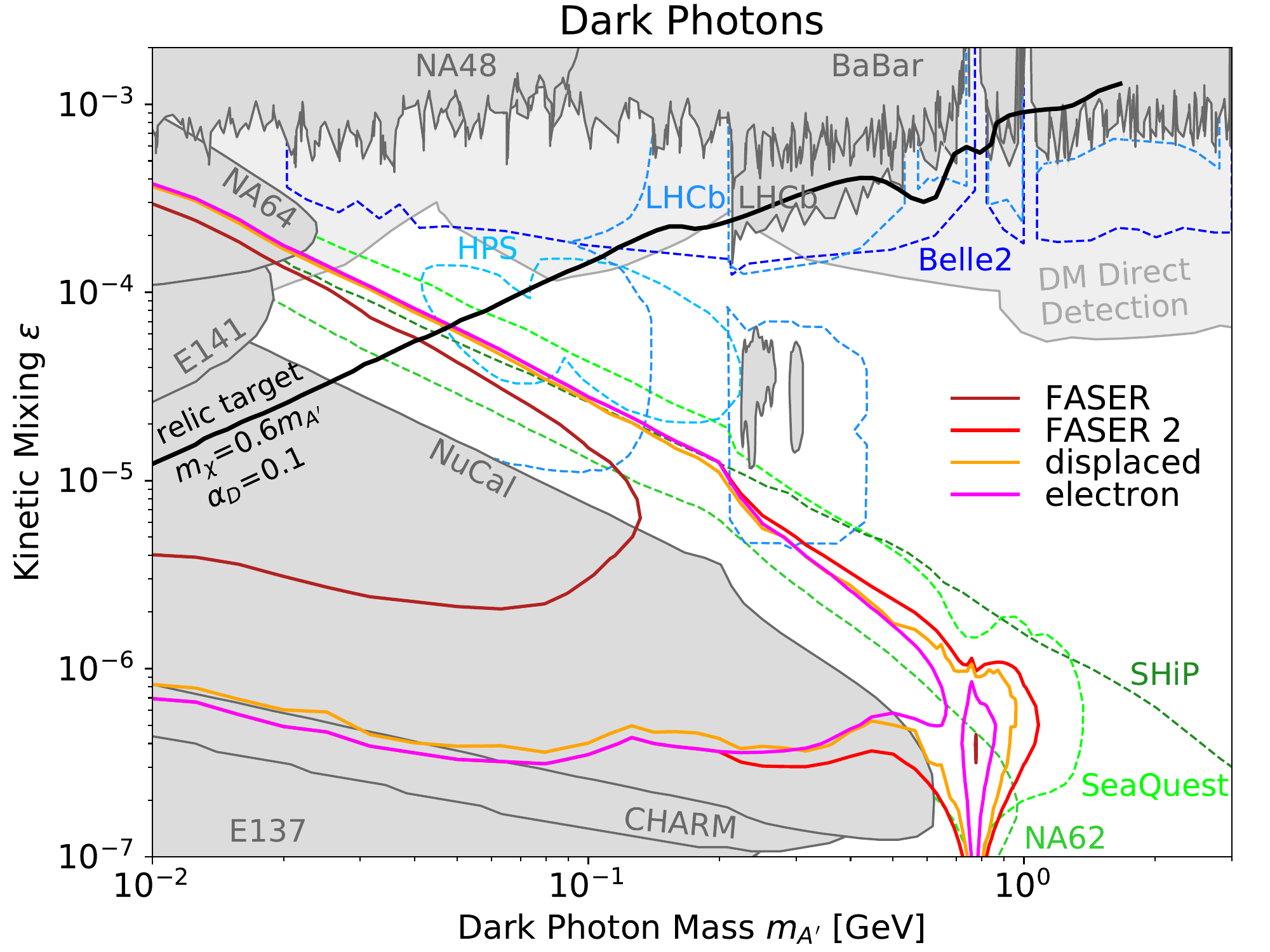}
\hfill
\includegraphics[width=0.49\textwidth]{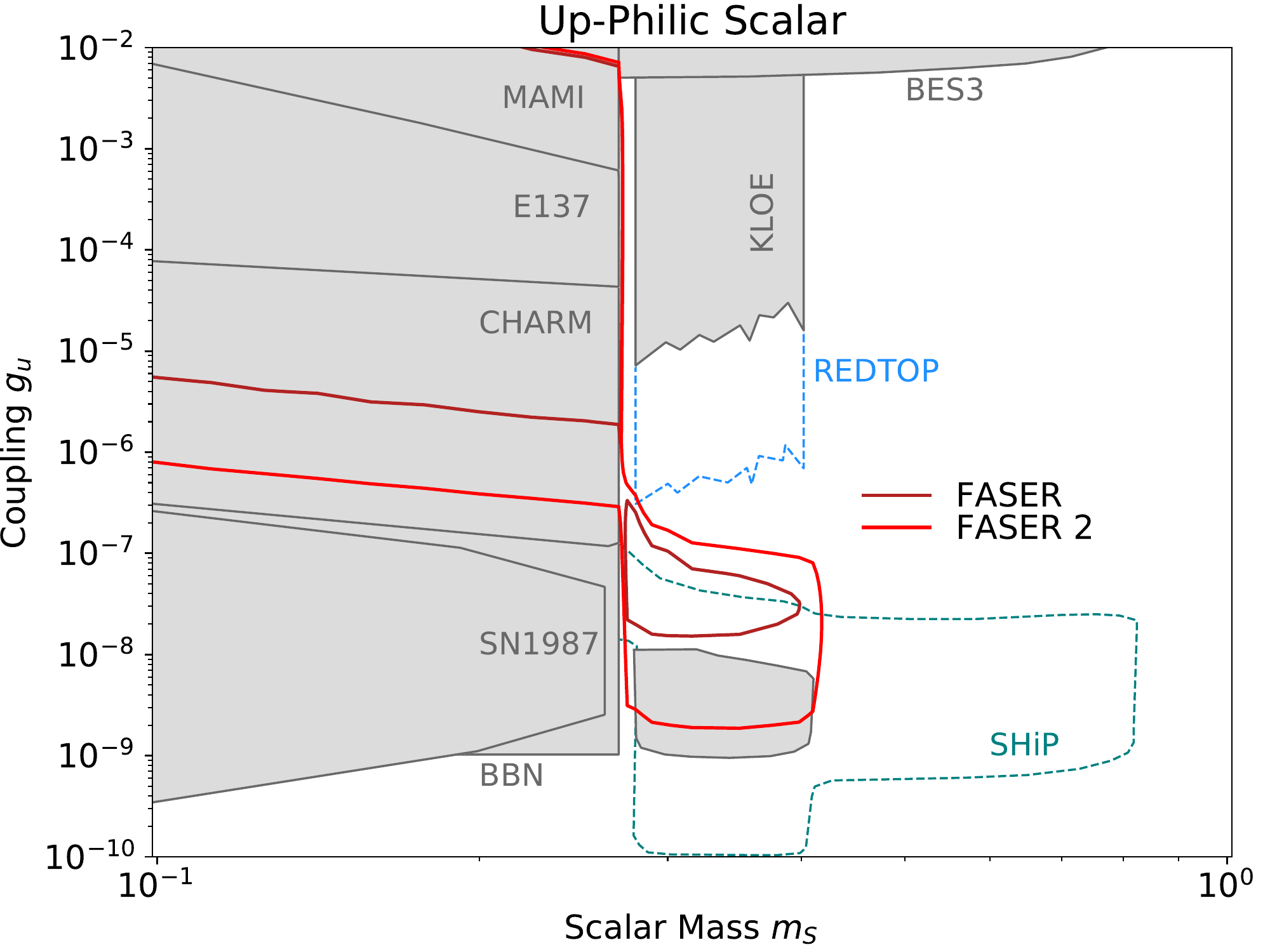}
\caption{Dark photon (left) and up-philic scalar (right) sensitivity reach lines in $(m_{A^\prime},\epsilon)$ and $(m_{S},g_u)$ planes, respectively, obtained for the FASER (brown solid lines) and FASER 2 (red) detectors. In the left panel, we also show for the dark photon model the expected FASER 2 sensitivity to only $A^\prime\to e^+e^-$ decay channel (magenta) and to all the possible visible decays but happening in a displaced detector by a $1~\m$ distance off the beam axis (orange). In both panels, we present previous bounds with dark gray-shaded regions and future sensitivity (colorful dashed lines) for selected other searches, as indicated in the plots. In the left panel, we also show with light-gray color current constraints on the dark photon parameter space from dark matter direct detection searches. In this case, we assume the complex scalar $\chi$ DM model with the fixed mass ratio $m_\chi/m_{A^\prime}=0.6$ and the dark coupling constant $\alpha_D=0.1$. See the text for more details and references for the bounds.}
\label{fig:LHC}
\end{figure*}

The sensitivity reach can also be degraded if the detector is not positioned along the beam collision axis, cf. Refs~\cite{Ariga:2018uku,Ariga:2018uku}, also Ref.~\cite{Boyarsky:2021moj} for the recent discussion about the SND@LHC experiment. However, this effect is not substantial as long as the displacement is not too large and the beam axis crosses the detector volume. In order to illustrate this, in the left panel of \cref{fig:LHC} we also present the corresponding sensitivity reach for the dark photon model of a FASER 2-like detector with the radius $R=1~\m$ assuming that its center has been shifted by $1~\m$ off the beam axis. The geometrical acceptance of the detector can be varied freely in \texttt{FORESEE}. 

We also show in the plot past bounds from the BaBaR~\cite{Lees:2014xha}, CHARM (following Ref.~\cite{Gninenko:2012eq}), E137~\cite{Bjorken:1988as}, E141~\cite{Riordan:1987aw}, LHCb~\cite{Aaij:2019bvg}, NA48/2~\cite{Batley:2015lha}, NA64~\cite{Banerjee:2018vgk}, and NuCal~\cite{Blumlein:1990ay} experiments, as well as complementary future sensitivity reach lines relevant for searches performed by Belle-II~\cite{Kou:2018nap}, HPS~\cite{Battaglieri:2017aum}, LHCb~\cite{Ilten:2015hya,Ilten:2016tkc}, NA62~\cite{Dobrich:2018ezn}, SeaQuest~\cite{Berlin:2018pwi}, and SHiP~\cite{Ahdida:2020new}, see Ref.~\cite{Beacham:2019nyx} for further discussion about future experiments searching for visibly decaying dark photons. 
\medskip 

One of the most important motivations for searching for LLPs is the role they can play as mediators between the SM and dark matter (DM) particles $\chi$. In particular, this allows for obtaining the correct value of the relic density, $\Omega_\chi^{\textrm{total}} h^2\simeq 0.12$~\cite{Aghanim:2018eyx}, via the popular thermal production mechanism~\cite{Boehm:2003hm,Pospelov:2007mp,Feng:2008ya}. Since such scenarios often predict both the mediator and DM masses to be in the sub-$\gev$ range, this could naturally lead to complementarity between searches for light mediator particles and those targeting DM direct detection (DD) signatures~\cite{Batell:2009di}.

In the left panel of \cref{fig:LHC}, we show the example thermal relic contour obtained for the scenario with the $A^\prime$ mediator and light complex scalar DM described by the following Lagrangian
\begin{equation}
\mathcal{L}_D \supset (D^\mu\chi)^\ast(D_\mu\chi) - m_\chi^2\chi^\ast\chi,
\end{equation}
where $D_\mu = \partial_\mu - ig_DA^{\prime}_\mu$ and we denote $\alpha_D = g_D^2/(4\pi)$. For the line presented in the plot, we fix the dark coupling constant at a value $\alpha_D=0.1$ and we assume that the mass ratio between the two dark species is always equal to $m_{\chi}/m_{A^\prime} = 0.6$ for varying dark photon mass. 

This choice of the mass ratio guarantees that the dark photon decays visibly into the SM species and can then be searched for in FASER and FASER 2 (see Ref.~\cite{Batell:2021blf} for the discussion about invisibly decaying forward-going dark photons). On the other hand, we require $\chi$ to be lighter than the dark photon. In this case, the DM relic density is set by annihilations into the SM fermions, $\chi\chi\to A^{\prime\ast}\to f\bar{f}$, and it is not suppressed by otherwise (if kinematically allowed) very efficient annihilations in the dark sector, $\chi\chi\to A^{\prime}A^{\prime}$. The particular value of the benchmark mass ratio used in \cref{fig:LHC} has been chosen to minimize the possible impact on $\Omega_\chi h^2$ of the forbidden DM annihilations  $\chi\chi\to A^\prime A^\prime$~\cite{DAgnolo:2015ujb}, the dark photon initial state radiation $\chi\chi\to A^\prime A^{\prime\ast}\to A^\prime \ell\ell$~\cite{Rizzo:2020jsm}, and the resonant annihilations into the SM particles, $\chi\chi\to A^\prime \to \ell\ell$, with $m_{A^\prime}\simeq 2 m_\chi$, cf. Refs~\cite{Feng:2017drg,Berlin:2020uwy,Bernreuther:2020koj} for recent discussion. Last but not least, we note that the annihilation cross section for scalar DM is $p$-wave suppressed, which allows one to avoid stringent bounds from the Cosmic Microwave Background (CMB) radiation~\cite{Slatyer:2009yq,Aghanim:2018eyx}.

As can be seen in the plot, a small FASER detector during Run 3 will already be able to probe an important part of the allowed region of the parameter space which corresponds to the correct value of the DM relic density. Notably, in this scenario, the FASER reach corresponds to the most cosmologically relevant region of the parameter space, in which $\chi$ DM is not overproduced in the early Universe. Instead, for the region below the thermal relic line, the correct DM abundance cannot be reproduced in this simple model without introducing further modifications in the allowed DM interactions or in the standard cosmological scenario. 

The role of far-forward search for $A^\prime$ in FASER at the LHC can be further emphasized by studying current constraints on such DM particles from DD searches in the underground detectors. In the left panel of \cref{fig:LHC}, we present with a light-gray shaded region current such dominant bounds obtained by the Xenon10~\cite{Angle:2011th} and Xenon1T~\cite{Aprile:2019xxb,Aprile:2019jmx} detectors. In the plot, we combine bounds from the DM-electron scatterings following Ref.~\cite{Essig:2017kqs}, as well from the searches for DM-nucleus interactions that employ the Migdal effect~\cite{Ibe:2017yqa}. When presenting these bounds, we also assume that $\Omega_\chi h^2 \simeq \Omega_\chi^{\textrm{total}}$ in the entire relevant region of the parameter space. 

After taking into account the DM DD searches, there remains only a narrow allowed region in the parameter space of the $A^\prime/\chi$ model under study, in which we do not expect DM to be thermally overproduced, i.e. we require $\Omega_\chi^{\textrm{thermal}} h^2\lesssim \Omega_\chi^{\textrm{total}}$. Notably, for $m_\chi\lesssim 20~\mev$ (i.e. $m_{A^\prime}\lesssim 40~\mev$) and $\epsilon\lesssim 10^{-4}$, this region overlaps with the reach of the FASER searches for visibly decaying $A^\prime$. The relevant detection prospects will be further strengthened in FASER 2 during HL-LHC. The corresponding DM particles can also be searched for in future DM DD experiments, see Ref.~\cite{Billard:2021uyg} for recent review. 

We stress that the DM DD bounds could be evaded e.g. for the inelastic DM scenario with the mass splitting between the two dark sector particles sufficient to suppress DM DD rates. Importantly, the presence of such a mass splitting between the two dark species would not affect the FASER sensitivity reach, unless it is large enough so that decays of the excited state can also contribute to the visible signal in the detector, cf. Refs~\cite{Berlin:2018jbm,Jodlowski:2019ycu} for further similar discussions. 

\subsection{Up-philic scalar}
\label{sec:upphilicscalar} 
 
We next consider the search for the light scalars coupled dominantly to the up quarks, cf. \cref{eq:Lupphilicscalar}. While typical LLP studies focus on the Yukawa-like interactions of BSM scalars that resemble the SM Higgs boson, coupling light scalars to only selected SM mass eigenstates leads to distinct phenomenology. In particular, by restricting the couplings to only up quarks, one can avoid stringent constraints from $B$ meson decays and other flavor bounds, while simultaneously keeping the production rate of such scalars in $pp$ collisions at a high level~\cite{Batell:2017kty,Batell:2018fqo}.

The dominant production modes of the up-philic scalars are in rare decays of $\eta$ and $\eta^\prime$ mesons, $\eta,\eta^\prime\to\pi^0 S$, while kaon and $B$ meson decays are suppressed in this case, differently from the light dark Higgs model. The scalars $S$ decay preferentially into hadronic final states, if kinematically allowed, with the dominant such decay channel into a pion pair. Instead, for lower scalar masses, the lifetime of $S$ is driven by loop-induced decays into photons and it, therefore, becomes significantly larger.

The drop in the lifetime of $S$ for $m_S$ growing above the pion threshold leads to specific features in the reach plot presented in the right panel of \cref{fig:LHC}. For $m_S<2m_{\pi^0}$, the expected sensitivity is limited by the large $S$ lifetime that allows LLPs to easily overshoot the detector. As a result, only values of the coupling constant above certain threshold, $g_u\gtrsim 10^{-5}$ ($10^{-6}$), can be probed in FASER (FASER 2). In addition, the relevant region of the parameter space has already been excluded by past searches in CHARM~\cite{Bergsma:1985qz}, E137~\cite{Bjorken:1988as}, and MAMI~\cite{Nefkens:2014zlt} experiments. For completeness, following Ref.~\cite{Batell:2018fqo}, we also show in the plot the constraints from the energy-loss rate in the core of SN1987A and from the number of effective degrees of freedom that could affect the Big Bang Nucleosynthesis (BBN).

\begin{figure*}[!t]
\centering
\includegraphics[width=0.49\textwidth]{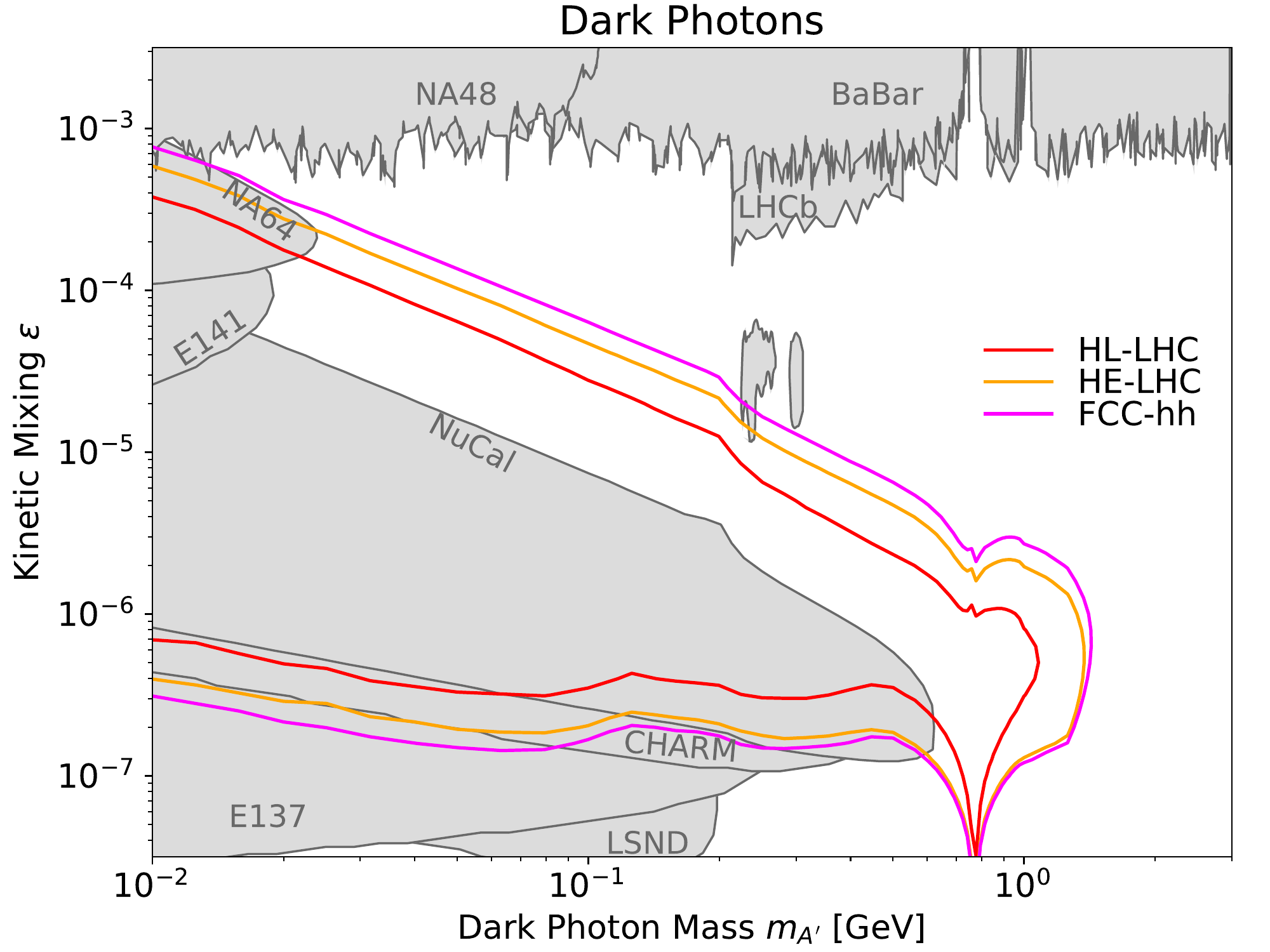}
\hfill
\includegraphics[width=0.49\textwidth]{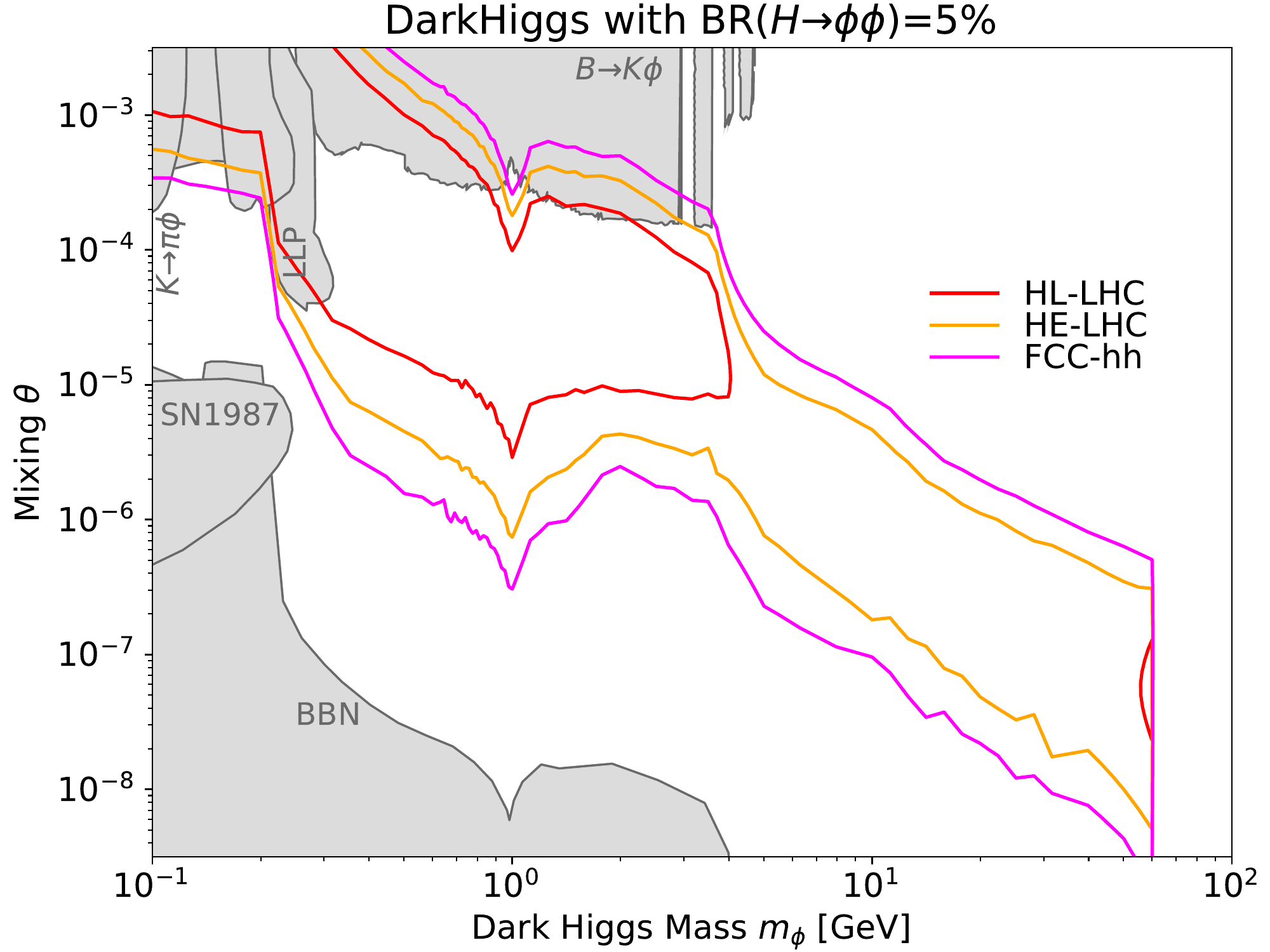}
\caption{Sensitivity reach lines obtained for the FASER 2 detector to take data during the HL-LHC era (red solid line) and for similar detectors operating at the future hadron colliders: HE-LHC (orange) and FCC-hh/SppC (purple). The details of the assumed detector design are given in \cref{tab:geometries}. The reach plots are shown for the dark photon model in a $(m_{A^\prime},\epsilon)$ plane (left) and for the dark Higgs boson model in a $(m_\phi,\theta)$ plane (right). In both cases, current bounds on the model parameter space are shown with gray-shaded regions (see the text for details).}
\label{fig:future}
\end{figure*}

On the other hand, for heavier $S$ with $m_S>2m_{\pi^0}$, we can take advantage of large LHC energies and boost factors to improve the previous bounds on more short-lived up-philic scalars. The corresponding expected sensitivity of both FASER and FASER 2 is, however, concentrated in the limited region of the parameter space. It corresponds to the dark scalar masses lying in the range $2\,m_{\pi} < m_S < m_\eta - m_{\pi}$. The far-forward searches at the LHC will allow one to probe the currently not excluded values of the dark coupling constant in between the lower CHARM bound and the upper constraints from the KLOE experiment~\cite{Anastasi:2016cdz}. Notably, the latter search was sensitive to $S\to \pi^+\pi^-$ decays and, therefore, it did not constrain dark scalar masses below the corresponding threshold, even though possible decays to lighter neutral pions would keep the $S$ lifetime relatively low in the remaining small mass window, $2\,m_{\pi^0} < m_S < 2\,m_{\pi^\pm}$. Instead, when estimating the FASER and FASER 2 reach, we take into account both decays into charged and neutral pions, with the latter leading to a striking $4\gamma$ signature in the detector. In the figure, we also show for comparison the expected future sensitivity of the proposed SHiP and REDTOP~\cite{Gatto:2016rae} experiments following Ref.~\cite{Batell:2018fqo}. 

\subsection{Future Colliders}
\label{sec:future} 

While we have so far focused on the LHC, the idea to search for LLP decay signatures in the far-forward region of the hadron colliders could also be employed in future such experimental facilities beyond the HL-LHC era. It is then useful to briefly analyze the relevant detection prospects in connection to the proposed such colliders, see, e.g.,  Ref.~\cite{Benedikt:2018ofy,Zimmermann:2018koi} for recent reviews. In particular, we will discuss the sensitivity reach of the detector similar in size and position to FASER 2, which could operate at the $27~\tev$ $pp$ collision energy in the era of the High-Energy LHC (HE-LHC)~\cite{Todesco:2011np}. We also present the sensitivity reach for $100~\tev$ energy. In this case, we choose for illustration a possible forward detector location based on the proposed concepts of the proton-proton Future Circular Collider (FCC-hh)~\cite{Mangano:2017tke} and the Super Proton-Proton Collider (SppC)~\cite{Gao:2017ssn}. We give the relevant parameters of the detectors in \cref{tab:geometries}.

While for the HE-LHC we employ the same location as for FASER and FASER 2, the $100~\tev$ colliders require shifting the detector to a larger distance $L$ from the IP. This is due to a smaller curvature of the beam pipe characteristic for larger colliders. In particular, to allow for sufficient shielding from the beam-induced background, we require the minimal distance between the far-forward detector and the (already curved) beam tunnel to be of order $5~\m$. We then expect that the minimal corresponding values of $L$ are equal to $1.1~\km$ for FCC-hh and $865~\m$ for SppC, respectively. In the following, for simplicity, we assume $L=1~\km$ for the detector operating at $100~\tev$ $pp$ collision energy. 

The relevant sensitivity reach lines for FASER 2 during HL-LHC, HE-LHC and FCC-hh are shown in \cref{fig:future} for the dark photon and dark Higgs models, cf. \cref{eq:LAprime,eq:darkHiggseffL}, respectively. As shown in the left panel, increasing the $pp$ collision energy and the boost factor of the produced LLPs will improve the expected sensitivity reach in the dark photon model. This is especially true for larger values of both $\epsilon$ and $m_{A^\prime}$ that corresponds to the smaller $A^\prime$ lifetime. It is worth stressing, though, that for the vanilla dark photon model this improvement is not significant. In fact, if fully exploited, the far-forward searches in FASER and FASER 2, along with other experimental probes, could almost saturate the relevant detection prospects already during the LHC and HL-LHC eras.

On the other hand, the detectors operating in both the HE-LHC and FCC-hh/SppC could have a more important impact on constraining the allowed parameter space of the dark Higgs model. We illustrate this in the right panel of \cref{fig:future}. In the plot, we also show with a gray-shaded region current bounds. The dominant ones come from the searches for dark scalars in kaon decays $K \to \pi \phi$ by NA62~\cite{Ruggiero:2020phq} and BNL-E949~\cite{Artamonov:2009sz}, B-meson decays $B \to K \phi$ by LHCb~\cite{Aaij:2015tna, Aaij:2016qsm}, and long-lived particle searches at CHARM~\cite{Winkler:2018qyg}, LSND~\cite{Foroughi-Abari:2020gju} and MicroBooNE~\cite{microboone}. We also show the limits from astrophysical observations of SN 1987A and the BBN bounds, following Ref.~\cite{Winkler:2018qyg}. 

For a light scalar $\phi$, which can be efficiently produced in rare $B$ meson decays, the search performed at $100~\tev$ energies will improve the sensitivity reach by an additional order of magnitude in the $\phi-h$ mixing angle $\theta$ in comparison with the FASER 2 experiment operating during HL-LHC. For the lowest values of $\theta$, the dominant production mode of $\phi$ in this region can be due to the $B$ meson decays into pairs of dark scalars via an off-shell Higgs boson, $b\to s h^\ast\to s\phi\phi$. This production mode remains relevant if the corresponding SM Higgs branching fraction is larger than $\mathcal{O}(0.1\%)$, as dictated by the trilinear coupling $\lambda$ in \cref{eq:darkHiggseffL}. This, however, only affects a narrow region in the covered parameter space, while the majority of the observed reach in this dark scalar mass range is not sensitive to $\lambda$.

The situation changes for dark scalar masses larger than the $B$ meson mass. Here, FASER 2 in the HL-LHC era will have almost no sensitivity reach, besides a very narrow region in the parameter space, in which $m_\phi\simeq m_h/2$~\cite{Boiarska:2019vid}. In this case, the dark scalars are pair-produced in decays of on-shell SM Higgs bosons, $h\to\phi\phi$. This production mode becomes even more important for similar such detectors operating at future colliders. The $pp$ collision energy then becomes large enough so that even the SM Higgs bosons with $m_h\simeq 125~\gev$ are sufficiently boosted to become relevant for far-forward searches. In particular, in the right panel of \cref{fig:future}, we assume $\textrm{BR}(h\to\phi\phi) = 5\%$ when obtaining the reach lines. In this case, the sensitivity reach could be extended to the values of $m_\phi$ spanning the entire mass range in between the $B$ meson mass and $m_\phi \simeq m_h/2$. This will also correspond to very low values of the mixing angle, $\theta\sim 10^{-9}$. Again, the impact of direct $\phi$ production in decays of the SM Higgs bosons will become much suppressed if the corresponding branching fraction falls below $0.1\%$.

\section{Conclusion}
\label{sec:conclusion} 

The upcoming LHC Run 3 will soon begin the next chapter in the continued efforts towards discovering signatures of new physics. Besides extending the existing experimental strategies, it will also initiate a new direction in BSM searches at the LHC in its far-forward region. In this case, when properly placed, even relatively small-size detectors can offer excellent discovery prospects. Notably, this has been recently demonstrated with the first observation of neutrino candidate events at the LHC~\cite{Abreu:2021hol}.

The most commonly discussed strategy to detect new forward-going light and long-lived particles produced abundantly at the $pp$ interaction point is via their displaced decays into SM species. These will be searched for in the distant FASER experiment. In the more distant future, a similar research agenda could be promoted to an even higher level by detectors taking data during the HL-LHC era that will be placed in the Forward Physics Facility.

This experimental program could much benefit from further phenomenological and theoretical analyses of new physics models that could be probed this way. In this study, we have introduced a numerical package \textbf{FOR}ward \textbf{E}xperiment \textbf{RE}ach \textbf{E}stimator, or \texttt{FORESEE}, that allows one to estimate the sensitivity reach of the far-forward experiments operating at $14$, $27$, or $100~\tev$ $pp$ collision energies in their searches for highly-displaced LLP decays. In addition, in the package, the validated far-forward spectra of light mesons and many other SM species have been made available to facilitate their use in separate studies, cf. \cref{sec:hadrons} for the discussion about the simulation details and the relevant references.

We have first illustrated the capabilities of the package by presenting the expected reach of the FASER and FASER 2 experiments in the dark photon and up-philic scalar models. In the first case, we have also discussed the possible complementarity between the FASER search for visibly decaying dark photons and DM direct detection experiments, assuming a simple model with complex scalar DM and the $A^\prime$ mediator. We stress that such a complementarity could best be seen for only a limited mass range of both dark species, $m_{A^\prime}/2 \lesssim m_\chi\lesssim m_{A^\prime}$. The cosmologically viable such scenarios are further constrained by requiring that the DM is not thermally overproduced in the early Universe. Interestingly, however, these conditions can be satisfied in the simple benchmark case precisely in the region of the parameter space of the model that will be covered by FASER during Run 3.

Last but not least, for the first time, we have also analyzed the discovery potential of possible far-forward experiments to operate in the future hadron colliders: HE-LHC, SppC, and FCC-hh. To this end, we have focused on the popular dark photon and dark Higgs boson models. We show that, while in the former case the capabilities of the far-forward searches can basically be saturated already during the HL-LHC phase, the latter BSM scenario could benefit from employing the increased $pp$ collision energy. This is, however, mainly due to possible decays of the on-shell Higgs bosons. The precise impact of employing the future colliders on the dark Higgs sensitivity reach will, therefore, depend on earlier measurements of the relevant invisible Higgs branching fraction.

\section*{Acknowledgements}

We thank Jason Arakawa, Jonathan Feng, Théo Moretti, Carlo Pandini and Lorenzo Paolozzi for useful discussions.  We are also grateful to the authors and maintainers of many open-source software packages, including
\texttt{CRMC}~\cite{CRMC},
\texttt{EPOS-LHC}~\cite{Pierog:2013ria},
\texttt{Jupyter} notebooks~\cite{soton403913}, 
\texttt{Matplotlib}~\cite{Hunter:2007}, 
\texttt{Pythia~8}~\cite{Sjostrand:2014zea},
\texttt{QGSJET-II-04}~\cite{Ostapchenko:2010vb},
\texttt{scikit-hep}~\cite{Rodrigues:2019nct}, 
\texttt{Sibyll~2.3c}~\cite{Ahn:2009wx,Riehn:2015oba}, and
\texttt{uproot}~\cite{jim_pivarski_2019_3256257}.
F.K. is supported by the Department of Energy under Grant No. DE-AC02-76SF00515. S.T. is supported by the project AstroCeNT, Particle Astrophysics Science and Technology Centre, carried out within the International Research Agendas programme of the Foundation for Polish Science financed by the European Union under the European Regional Development Fund. S.T. is also supported in part by the Polish Ministry of Science and Higher Education through its scholarship for young and outstanding scientists (decision no 1190/E-78/STYP/14/2019).



\bibliography{references}

\end{document}